\documentclass[nofootinbib,aps,preprint,longbibliography]{revtex4-2}
\pdfoutput=1

\usepackage{gensymb}
\usepackage{color}
\usepackage{graphicx}
\usepackage{epstopdf}
\usepackage{dcolumn}
\usepackage{bm}
\usepackage{amsmath}

\newcommand{\be}{\begin{equation}}
\newcommand{\ee}{\end{equation}}
\newcommand{\bfig}{\begin{figure}}
\newcommand{\efig}{\end{figure}}

\begin{document}
\title{Kagomé lattice promotes chiral spin fluctuations}
\author{Kamil K. Kolincio$^{1,2,\ddagger}$}\email{kamil.kolincio@pg.edu.pl}
\thanks{$\ddagger$ These two authors contributed equally to this work}
\author{Max Hirschberger$^{1,3,\ddagger}$}\email{hirschberger@ap.t.u-tokyo.ac.jp}
\author{Jan Masell$^{1,4}$}
\author{Taka-hisa Arima$^{1,5}$}
\author{Naoto Nagaosa$^{1,3}$}
\author{Yoshinori Tokura$^{1,3,6}$}
\affiliation{$^{1}$RIKEN Center for Emergent Matter Science (CEMS), Wako, Saitama 351-0198, Japan}
\affiliation{$^{2}$Faculty of Applied Physics and Mathematics, Gdańsk University of Technology, Narutowicza 11/12, 80-233 Gdańsk, Poland}
\affiliation{$^{3}$Department of Applied Physics and Quantum-Phase Electronics Center (QPEC), The University of Tokyo, Bunkyo-ku, Tokyo 113-8656, Japan}
\affiliation{$^{4}$Institute of Theoretical Solid State Physics, Karlsruhe Institute of Technology (KIT), 76049 Karlsruhe, Germany}
\affiliation{$^{5}$Department of Advanced Materials Science, The University of Tokyo, Kashiwa 277-8561, Japan}
\affiliation{$^{6}$Tokyo College, The University of Tokyo, Bunkyo-ku, Tokyo 113-8656, Japan}

\begin{abstract}
Magnetic materials with tilted electron spins often exhibit conducting behaviour that cannot be explained from semiclassical theories without invoking fictitious (emergent) electromagnetic fields. Quantum-mechanical models explaining such phenomena are rooted in the concept of a moving quasiparticle’s Berry phase\cite{1,2}, driven by a chiral (left- or right-handed) spin-habit. Dynamical and nearly random spin fluctuations, with a slight bent towards left- or right-handed chirality, represent a promising route to realizing Berry-phase phenomena at elevated temperatures\cite{3,4,5,6}, but little is known about the effect of crystal lattice geometry on the resulting macroscopic observables. Here, we report thermoelectric and electric transport experiments on two metals with large magnetic moments on a triangular and on a slightly distorted kagomé lattice, respectively. We show that the impact of chiral spin fluctuations is strongly enhanced for the kagomé lattice. Both these spiral magnets have similar magnetic phase diagrams including a periodic array of magnetic skyrmions. However, our modelling shows that the geometry of the kagomé lattice, with corner-sharing spin-trimers, helps to avoid cancellation of Berry-phase contributions; spin fluctuations are endowed with a net chiral habit already in the thermally disordered (paramagnetic) state. Hence, our observations for the kagomé material contrast with theoretical models treating magnetization as a continuous field\cite{7,8,9,10,11}, and emphasize the role of lattice geometry on emergent electrodynamic phenomena.
\end{abstract}
\maketitle  

Spin dynamical processes, where chiral magnetic textures appear in a transient fashion, have become experimentally accessible to modern magnetism research with the rise of ultrafast optical and x-ray techniques\cite{12,13,14}. Recent theories emphasize rectification processes facilitated by transient spin arrangements: during their rapid movement on the time-scale of picoseconds, clusters of thermally agitated spins on lattice sites $i$, $j$, $k$ are endowed with a local scalar spin chirality, mathematically described as $\chi=\mathbf{S}_i\cdot (\mathbf{S}_j\times \mathbf{S}_k)$, whose thermal average $\left<\chi\right>_T$, referred to by the shorthand SSC in the following, can be constant in time despite absence of long-range order \cite{4,5,15,16}. The development of design principles for materials with fluctuation-driven SSC, particularly as related to exploitation of common lattice motifs, promises new impulses for controlling heat and charge transport~\cite{4,5,15,17}, magneto-optical responses~\cite{18,19}, and the second harmonic generation of light~\cite{20}. Moving beyond previous experimental reports focused on the phenomenological observation of this effect~\cite{15,17,21,22,23}, we provide an expanded view of spin chirality driven by thermal fluctuations, emphasizing the atomistic lattice as a key factor in its amplification or suppression.

To investigate the role of lattice geometry in SSC from thermal fluctuations, we have chosen two reference materials: Gd$_3$Ru$_4$Al$_{12}$ and Gd$_2$PdSi$_3$, metallic magnets with Néel temperatures of $18.6$ and $21\,$K respectively~\cite{24,25}, known to host skyrmion lattices~\cite{26,27}. Both materials crystallize in hexagonal structures~\cite{25,28,29}, and the main building blocks of their magnetic sublattices are triangles of trivalent Gadolinium ions: Fig. 1 (a,c). The spatial arrangement of these basic triangular units, however, constitutes a major difference between them: Gd$_2$PdSi$_3$’s triangular lattice is a plain array of closely packed, edge-sharing, and identical triangles. Meanwhile, the magnetic moments in Gd$_3$Ru$_4$Al$_{12}$ are arranged on a ‘breathing’ kagomé lattice, where alternating large and small triangles share corners with each other, being interspersed with hexagons~\cite{30}. The kagomé lattice is known for harbouring a number of novel magnetic phenomena, ranging from spin-liquid formation in frustrated antiferromagnets~\cite{31} to chiral spin states~\cite{32} and topological spin-waves in ferromagnets~\cite{33,34}. As demonstrated here, the kagomé net is also a minimal model system for the exploration of spin chirality in thermal fluctuation processes. 

\textbf{Spin chirality on the breathing kagomé lattice}

We illustrate the key point of our study in Fig. 1, focusing on thermal fluctuation processes that can generate a net chiral habit in solids. In the high temperature limit, depicted in Fig. 1 (a), spins fluctuate wildly in all directions of space. In this entirely random state, it is equally likely for a given spin cluster to carry positive or negative local chirality, even when a magnetic field is applied to break time-reversal symmetry. Upon cooling, short-range correlations develop and, on average, a single triangular unit may acquire finite chirality; however, depending on the lattice type, this SSC is amplified or cancelled out, as demonstrated in the present work using a kagomé lattice and a triangular lattice, respectively [Fig. 1(b,c)].

Unbiased numerical calculations in the high-temperature state, detailed in Methods, yield almost equal probability $p(\chi)$ for negative and positive spin chirality on a given triangle, corresponding to a symmetric curve in the inset of Fig. 1(d), or to vanishingly small net probability $p_\mathrm{net}(\chi)=p(\chi)-p(-\chi)$  in the main panel. The total sideways force on (blue) conduction electrons in Fig. 1(a,d) is $\sim \int_{-\infty}^\infty d\chi p(\chi)\chi\sim \int_0^\infty d\chi \chi p_{net} (\chi)$. Hence, although the instantaneous lattice state represented in Fig. 1(a) may have an imbalance of triangles with positive and negative local chirality, there is no averaged SSC and no impact on conduction electron motion. As thermal agitation decreases and short-range spin-spin interactions become pronounced, we approach the transition to long-range order: in this regime, even a weak external magnetic field causes unbalancing of $p(\chi)$, leading to finite SSC as described by the asymmetric probability distribution in Fig. 1(e). Both kagomé and triangular lattices are composed of triangle spin clusters, suitably described by the calculations in Fig. 1(d,e); however, equivalent neighboring triangle plaquettes, shown in Fig. 1(c), yield cancelling spin chirality for the triangular lattice. 

Conduction electrons feel the SSC as an emergent magnetic field, and are deflected transverse to the direction of the applied current~\cite{2}. Experimentally, anomalous electron velocities generated by SSC can be directly probed in transport experiments, as they contribute to Hall and Nernst effects of solids. The initial question to be answered here is, hence, whether and how thermal fluctuations with SSC appear in basic transport properties of a kagomé and a triangular lattice. Figure 2(a-b) compare the SSC-induced contribution to the Hall conductivity for the kagomé system Gd$_3$Ru$_4$Al$_{12}$; the separation of contributions to the Hall and Nernst effects is discussed in Methods. Starting from zero-field, the Hall effect rapidly increases: Even as spins are partially aligned, they retain the freedom to fluctuate around their average direction and to realize spin-chiral configurations. After attaining a maximum at intermediate fields, the Hall data drop when all spins are forcibly aligned. This behaviour can be summarized in terms of a generic scaling between magnetization and the chiral part of the Hall conductivity, provided in Extended Figure 1.

Contour maps in Fig. 2(b,d) show that lattice-averaged spin chirality is present already in the paramagnetic regime, growing larger and larger when approaching from above the boundary of the magnetically ordered phase. The experimental profile of the Hall effect in the plane of temperature and magnetic field is in good qualitative agreement with our numerical calculations, as discussed further in Extended Note 1 and Extended Figure 2(a), and brings to mind early theoretical work~\cite{3,4,5}. This kind of plume-shaped contour emerges because increasingly higher fields are required to align magnetic moments and to suppress SSC as the sample is heated. The signal finally disappears at elevated temperatures, where thermal motion is so rapid – and short-range correlations are so weak – that the imbalance of positive and negative local chirality becomes negligible. Note that the SSC-driven signal can also be made to vanish at low temperature, when all magnetic moments are forcibly coaligned by large external magnetic fields. 

\textbf{Cancellation of spin chirality on the triangular lattice}

We proceed to contrast data from the kagomé material with the simpler triangular lattice geometry in Gd$_2$PdSi$_3$, for which Fig. 2(c-d) shows the chirality-induced Hall effect. The Hall signal in the paramagnetic regime is much smaller than its counterpart in the long-range ordered skyrmion lattice phase~\cite{26}, and – we emphasize – at least one order of magnitude smaller as compared to the kagomé material Gd$_3$Ru$_4$Al$_{12}$. We propose that a significant suppression of the fluctuation-induced SSC is rooted in the geometry of the triangular network: Here, each spin is shared between six equivalent triangles. While the weak canting of three next-neighbour spins produces local spin chirality for an individual triad, contributions of neighbouring triangles tend to have opposite sign, as illustrated in Fig. 1(c). The spin chirality cancels to zero when integrated over the entire magnetic plane, under the condition that the solid angle covered by fluctuations is small [Fig. 2(e), left]. Large-amplitude spin fluctuations however, e.g. slowly fluctuating topological defects such as topological $Z_2$ vortices, skyrmions and hedgehogs~\cite{35,36,37,38}, escape from this extinction principle [Fig. 2(e), right]. A small, yet finite chirality signal from thermal agitation can hence be created even on a simple lattice such as the triangular net of Gd$_2$PdSi$_3$; see Extended Note 1 for the possible role of large-angle fluctuations.

Our numerical model for thermal fluctuations on the kagomé lattice is further supported by the observation of a power-law decay for the Hall signal at constant field, shown in Fig. 2(f). We find an inverse quadratic decay of the Hall signal with temperature, well-described by an analytic high-temperature expansion of the spin chirality (Extended Note 1). The theoretical model takes into account chiral Dzyaloshinskii-Moriya interactions (DMI) on individual spin-triads of the kagomé network. In particular, the observed exponent $2$ rules out SSC driven by anisotropic exchange interactions, by single-ion compass anisotropy, or by dynamically-fluctuating, long-period spin textures, such as fluctuating skyrmion spin-vortices. For all of these mechanisms, larger exponents are expected – see also Ref. \cite{4}. 

For comparison of triangular and kagomé lattices, we carefully chose samples with comparable metallicity, as illustrated in Fig. 3(a). Analysis of fluctuation-driven transport phenomena is facilitated in less conducting crystals by a suppression of backgrounds from the normal Hall effect as well as extrinsic skew-scattering~\cite{39} (Methods and Extended Fig. 4). 

\textbf{Traces of spin chirality in the thermoelectric Nernst response}

We proceed by comparing electric and thermoelectric transport properties. The experimental geometry for the presently reported thermoelectric Nernst measurement is indicated in Fig. 3(b), inset: A transverse electric current $J_y$ is driven by the applied temperature gradient $(-\nabla T)$, analogous to Hall’s electric current driven by and applied electric field. The thermoelectric Nernst conductivity $\alpha_{xy} \sim J_y / (-\nabla T)$ includes a contribution from spin chirality, which is directly apparent in the ratio $\Gamma = \left.\alpha_{xy}/(TM)\right|_{B\rightarrow +0}$: Figure 3(b) shows nearly temperature independent behaviour of this ratio for both compounds between room temperature and approximately $80\,$Kelvin, related to the anomalous Nernst effect from spin-orbit coupling (Methods). Below $80\,$Kelvin, a broad depression of said ratio appears only for the kagomé compound. This regime, where the Nernst conductivity does not follow the magnetization even in the absence of long-range magnetic order, marks the emergence of chiral spin fluctuations above the transition to long-range order. In contrast, there is no clear Nernst anomaly for the triangular lattice, evidencing the suppression of SSC in this simpler lattice type. Figure 3(b) further demonstrates sharply diverging data traces inside the magnetically ordered phases, which are discussed in Extended Fig. 3. Our contour map of the SSC-driven Nernst signal in Fig. 3(c) is consistent with the Hall effect of Fig. 2(c), where in both cases the signal appears at the boundary of the ordered phase. The maximum amplitude of Hall and Nernst conductivities shifts slowly towards elevated magnetic fields when increasing temperature.

\textbf{Semi-quantitative analysis}

The present, centrosymmetric intermetallics are representative members of an emerging materials family with an ultra-dense lattice of skyrmion spin-vortex tubes (skyrmion diameter $\sim 2-3\,$nanometres)~\cite{26,27}. In the skyrmion lattice, atomic spins describe a $4\pi$ winding within each magnetic unit cell, hence providing a natural benchmark for the SSC generated by thermal fluctuations~\cite{40}. The skyrmion lattice~\cite{26} produces a 1.4 times larger Hall effect than the maximal value attained due to thermal fluctuations [Fig. 4(a), top]. As one magnetic skyrmion in Gd$_3$Ru$_4$Al$_{12}$ covers roughly twenty-seven small triangles of the kagomé network, our calculation in Methods predicts 1.2 times as large as spin chirality in the skyrmion lattice as compared to the fluctuating regime. The reasonable agreement between experiment and this simple estimate justifies the choice of model parameters in the numerical calculations. Meanwhile, the experimental data for Gd$_2$PdSi$_3$ suggest that SSC in the fluctuating regime is not only significantly suppressed in comparison with Gd$_3$Ru$_4$Al$_{12}$, but also about one order of magnitude weaker than Hall and Nernst features observed in the skyrmion phase of this compound~\cite{27,41} [Fig. 4(a), bottom].

The thermoelectric ratio $\alpha_{xy}^\chi / \sigma_{xy}^\chi$ of SSC-related Hall and Nernst conductivities [Fig. 4(a,b)] indicates how dramatically the emergent magnetic fields depend on a change of the chemical potential. In presence of singularities in the electronic structure of solids, e.g. in presence of Weyl-type monopoles or line degeneracies, the thermoelectric ratio is expected to approach the quantized value $k_B/e$ when the thermal energy $k_B T$ is larger than disorder-induced level broadening $\hbar/\tau$ with carrier lifetime $\tau$~\cite{42,43, 44, 17}. We report a notable suppression of the thermoelectric ratio to $0.04$ in the skyrmion lattice phases of both compounds, where the heat bath temperature is far below $\hbar/\tau\sim 70\,$Kelvin, estimated in Extended Note 3. Meanwhile, we find that the thermoelectric ratio in the fluctuating regime is on the order of $0.1\,k_B/e$ – not far from unity and consistent with a complex interplay between thermal spin fluctuations, singularities in the electronic band structure, and the quantum-mechanical Berry phase of conduction electrons.

\textbf{Conclusions and perspective}

We conclude with a broader classification of two-dimensional magnetic lattices and their role in fluctuation phenomena. Starting with lattices made from equilateral polygons, we define trivial tiling motifs to be those where cancellation of local spin chirality occurs due to identical neighbouring polygons. Of these, there are four basic types: square, triangular, and honeycomb lattices, assembled from regular polygons, as well as the rhombic tiling. Tiling motifs of the second group, labelled as non-trivial in Fig. 4(c), avoid cancellation due to the presence of at least two types of polygons. Among these, there are precisely two structures which have only one type of nearest-neighbour bond: The kagomé lattice, and a tiling of two rhombi with dissimilar angles. For example, the kagomé lattice avoids cancellation of SSC by interspersing triangles with hexagons. Further relaxing the condition of equivalent bonds, a vast number of non-trivial, two-dimensional tiling motifs can be considered and realized in solids~\cite{45,46}, for example the present breathing kagomé network of Gd$_3$Ru$_4$Al$_{12}$. 

We note that the implications of our work are not restricted to two dimensional, periodic tiling motifs: Firstly, quasi-layered sublattices of three-dimensional materials are amenable to such discussions. For example, the pyrochlore lattice of corner-sharing tetrahedra, a famous platform for frustration physics, features kagomé layers when viewed along the cubic $\left<111\right>$ direction. Secondly, quasi-crystalline magnets, e.g. a Penrose tiling of rhombi, are classified as nontrivial in our scheme and are hence expected to generate large fluctuation-induced effects related to spin chirality.

\textit{Acknowledgments.} J.M. was supported as Humboldt/JSPS International Research Fellow (19F19815) and by the Alexander von Humboldt Foundation as a Feodor Lynen Return Fellow. M.H. benefited from JSPS KAKENHI Grants No. JP21K13877 and JP22H04463, while also acknowledging a grant by the Fujimori Science and Technology Foundation. This work was partially sponsored by Core Research for Evolutional Science and Technology (CREST) Grant Nos. JPMJCR1874 and JPMJCR20T1 (Japan) from the Japan Science and Technology Agency. 

\bibliography{Gd3Ru4Al12_fluctuations}

\begin{thebibliography}{46}%
\makeatletter
\providecommand \@ifxundefined [1]{%
 \@ifx{#1\undefined}
}%
\providecommand \@ifnum [1]{%
 \ifnum #1\expandafter \@firstoftwo
 \else \expandafter \@secondoftwo
 \fi
}%
\providecommand \@ifx [1]{%
 \ifx #1\expandafter \@firstoftwo
 \else \expandafter \@secondoftwo
 \fi
}%
\providecommand \natexlab [1]{#1}%
\providecommand \enquote  [1]{``#1''}%
\providecommand \bibnamefont  [1]{#1}%
\providecommand \bibfnamefont [1]{#1}%
\providecommand \citenamefont [1]{#1}%
\providecommand \href@noop [0]{\@secondoftwo}%
\providecommand \href [0]{\begingroup \@sanitize@url \@href}%
\providecommand \@href[1]{\@@startlink{#1}\@@href}%
\providecommand \@@href[1]{\endgroup#1\@@endlink}%
\providecommand \@sanitize@url [0]{\catcode `\\12\catcode `\$12\catcode
  `\&12\catcode `\#12\catcode `\^12\catcode `\_12\catcode `\%12\relax}%
\providecommand \@@startlink[1]{}%
\providecommand \@@endlink[0]{}%
\providecommand \url  [0]{\begingroup\@sanitize@url \@url }%
\providecommand \@url [1]{\endgroup\@href {#1}{\urlprefix }}%
\providecommand \urlprefix  [0]{URL }%
\providecommand \Eprint [0]{\href }%
\providecommand \doibase [0]{https://doi.org/}%
\providecommand \selectlanguage [0]{\@gobble}%
\providecommand \bibinfo  [0]{\@secondoftwo}%
\providecommand \bibfield  [0]{\@secondoftwo}%
\providecommand \translation [1]{[#1]}%
\providecommand \BibitemOpen [0]{}%
\providecommand \bibitemStop [0]{}%
\providecommand \bibitemNoStop [0]{.\EOS\space}%
\providecommand \EOS [0]{\spacefactor3000\relax}%
\providecommand \BibitemShut  [1]{\csname bibitem#1\endcsname}%
\let\auto@bib@innerbib\@empty
\bibitem [{\citenamefont {Berry}(1984)}]{1}%
  \BibitemOpen
  \bibfield  {author} {\bibinfo {author} {\bibfnamefont {M.~V.}\ \bibnamefont
  {Berry}},\ }\bibfield  {title} {\bibinfo {title} {{Quantal phase factors
  accompanying adiabatic changes}},\ }\href@noop {} {\bibfield  {journal}
  {\bibinfo  {journal} {Proc R Soc Lond A}\ }\textbf {\bibinfo {volume}
  {392}},\ \bibinfo {pages} {45–57} (\bibinfo {year} {1984})}\BibitemShut
  {NoStop}%
\bibitem [{\citenamefont {Bruno}\ \emph {et~al.}(2004)\citenamefont {Bruno},
  \citenamefont {Dugaev},\ and\ \citenamefont {Taillefumier}}]{2}%
  \BibitemOpen
  \bibfield  {author} {\bibinfo {author} {\bibfnamefont {P.}~\bibnamefont
  {Bruno}}, \bibinfo {author} {\bibfnamefont {V.}~\bibnamefont {Dugaev}},\ and\
  \bibinfo {author} {\bibfnamefont {M.}~\bibnamefont {Taillefumier}},\
  }\bibfield  {title} {\bibinfo {title} {Topological {H}all effect and {B}erry
  phase in magnetic nanostructures},\ }\href
  {https://doi.org/10.1103/PhysRevLett.93.096806} {\bibfield  {journal}
  {\bibinfo  {journal} {Phys. Rev. Lett.}\ }\textbf {\bibinfo {volume} {93}},\
  \bibinfo {pages} {096806} (\bibinfo {year} {2004})}\BibitemShut {NoStop}%
\bibitem [{\citenamefont {Rózsa}\ \emph {et~al.}(2016)\citenamefont {Rózsa},
  \citenamefont {Simon}, \citenamefont {Palotás}, \citenamefont {Udvardi},\
  and\ \citenamefont {Szunyogh}}]{3}%
  \BibitemOpen
  \bibfield  {author} {\bibinfo {author} {\bibfnamefont {L.}~\bibnamefont
  {Rózsa}}, \bibinfo {author} {\bibfnamefont {E.}~\bibnamefont {Simon}},
  \bibinfo {author} {\bibfnamefont {K.}~\bibnamefont {Palotás}}, \bibinfo
  {author} {\bibfnamefont {L.}~\bibnamefont {Udvardi}},\ and\ \bibinfo {author}
  {\bibfnamefont {L.}~\bibnamefont {Szunyogh}},\ }\bibfield  {title} {\bibinfo
  {title} {{Complex magnetic phase diagram and skyrmion lifetime in an
  ultrathin film from atomistic simulations}},\ }\href@noop {} {\bibfield
  {journal} {\bibinfo  {journal} {Phys. Rev. B}\ }\textbf {\bibinfo {volume}
  {93}},\ \bibinfo {pages} {024417} (\bibinfo {year} {2016})}\BibitemShut
  {NoStop}%
\bibitem [{\citenamefont {Hou}\ \emph {et~al.}(2017)\citenamefont {Hou},
  \citenamefont {Yu}, \citenamefont {Daly},\ and\ \citenamefont {Zang}}]{4}%
  \BibitemOpen
  \bibfield  {author} {\bibinfo {author} {\bibfnamefont {W.-T.}\ \bibnamefont
  {Hou}}, \bibinfo {author} {\bibfnamefont {J.-X.}\ \bibnamefont {Yu}},
  \bibinfo {author} {\bibfnamefont {M.}~\bibnamefont {Daly}},\ and\ \bibinfo
  {author} {\bibfnamefont {J.}~\bibnamefont {Zang}},\ }\bibfield  {title}
  {\bibinfo {title} {{Thermally driven topology in chiral magnets}},\
  }\href@noop {} {\bibfield  {journal} {\bibinfo  {journal} {Phys. Rev. B}\
  }\textbf {\bibinfo {volume} {96}},\ \bibinfo {pages} {140403} (\bibinfo
  {year} {2017})}\BibitemShut {NoStop}%
\bibitem [{\citenamefont {Böttcher}\ \emph {et~al.}(2018)\citenamefont
  {Böttcher}, \citenamefont {Heinze}, \citenamefont {Egorov}, \citenamefont
  {Sinova},\ and\ \citenamefont {Dupé}}]{5}%
  \BibitemOpen
  \bibfield  {author} {\bibinfo {author} {\bibfnamefont {M.}~\bibnamefont
  {Böttcher}}, \bibinfo {author} {\bibfnamefont {S.}~\bibnamefont {Heinze}},
  \bibinfo {author} {\bibfnamefont {S.}~\bibnamefont {Egorov}}, \bibinfo
  {author} {\bibfnamefont {J.}~\bibnamefont {Sinova}},\ and\ \bibinfo {author}
  {\bibfnamefont {B.}~\bibnamefont {Dupé}},\ }\bibfield  {title} {\bibinfo
  {title} {{B–T phase diagram of Pd/Fe/Ir(111) computed with parallel
  tempering Monte Carlo}},\ }\href@noop {} {\bibfield  {journal} {\bibinfo
  {journal} {New J. Phys.}\ }\textbf {\bibinfo {volume} {20}},\ \bibinfo
  {pages} {103014} (\bibinfo {year} {2018})}\BibitemShut {NoStop}%
\bibitem [{\citenamefont {Kato}\ and\ \citenamefont {Ishizuka}(2019)}]{6}%
  \BibitemOpen
  \bibfield  {author} {\bibinfo {author} {\bibfnamefont {Y.}~\bibnamefont
  {Kato}}\ and\ \bibinfo {author} {\bibfnamefont {H.}~\bibnamefont
  {Ishizuka}},\ }\bibfield  {title} {\bibinfo {title} {{Colossal Enhancement of
  Spin-Chirality-Related Hall Effect by Thermal Fluctuation}},\ }\href@noop {}
  {\bibfield  {journal} {\bibinfo  {journal} {Phys. Rev. Appl.}\ }\textbf
  {\bibinfo {volume} {12}},\ \bibinfo {pages} {021001} (\bibinfo {year}
  {2019})}\BibitemShut {NoStop}%
\bibitem [{\citenamefont {Neubauer}\ \emph {et~al.}(2009)\citenamefont
  {Neubauer}, \citenamefont {Pfleiderer}, \citenamefont {Binz}, \citenamefont
  {Rosch}, \citenamefont {Ritz}, \citenamefont {Niklowitz},\ and\ \citenamefont
  {B{\"o}ni}}]{7}%
  \BibitemOpen
  \bibfield  {author} {\bibinfo {author} {\bibfnamefont {A.}~\bibnamefont
  {Neubauer}}, \bibinfo {author} {\bibfnamefont {C.}~\bibnamefont
  {Pfleiderer}}, \bibinfo {author} {\bibfnamefont {B.}~\bibnamefont {Binz}},
  \bibinfo {author} {\bibfnamefont {A.}~\bibnamefont {Rosch}}, \bibinfo
  {author} {\bibfnamefont {R.}~\bibnamefont {Ritz}}, \bibinfo {author}
  {\bibfnamefont {P.}~\bibnamefont {Niklowitz}},\ and\ \bibinfo {author}
  {\bibfnamefont {P.}~\bibnamefont {B{\"o}ni}},\ }\bibfield  {title} {\bibinfo
  {title} {Topological {H}all effect in the ${A}$ phase of {M}n{S}i},\
  }\href@noop {} {\bibfield  {journal} {\bibinfo  {journal} {Phys. Rev.
  Letters}\ }\textbf {\bibinfo {volume} {102}},\ \bibinfo {pages} {186602}
  (\bibinfo {year} {2009})}\BibitemShut {NoStop}%
\bibitem [{\citenamefont {Binz}\ and\ \citenamefont {Vishwanath}(2006)}]{8}%
  \BibitemOpen
  \bibfield  {author} {\bibinfo {author} {\bibfnamefont {B.}~\bibnamefont
  {Binz}}\ and\ \bibinfo {author} {\bibfnamefont {A.}~\bibnamefont
  {Vishwanath}},\ }\bibfield  {title} {\bibinfo {title} {{Theory of helical
  spin crystals: Phases, textures, and properties}},\ }\href@noop {} {\bibfield
   {journal} {\bibinfo  {journal} {Phys. Rev. B}\ }\textbf {\bibinfo {volume}
  {74}},\ \bibinfo {pages} {214408} (\bibinfo {year} {2006})}\BibitemShut
  {NoStop}%
\bibitem [{\citenamefont {Binz}\ \emph {et~al.}(2006)\citenamefont {Binz},
  \citenamefont {Vishwanath},\ and\ \citenamefont {Aji}}]{9}%
  \BibitemOpen
  \bibfield  {author} {\bibinfo {author} {\bibfnamefont {B.}~\bibnamefont
  {Binz}}, \bibinfo {author} {\bibfnamefont {A.}~\bibnamefont {Vishwanath}},\
  and\ \bibinfo {author} {\bibfnamefont {V.}~\bibnamefont {Aji}},\ }\bibfield
  {title} {\bibinfo {title} {{Theory of the Helical Spin Crystal: A Candidate
  for the Partially Ordered State of MnSi}},\ }\href@noop {} {\bibfield
  {journal} {\bibinfo  {journal} {Phys. Rev. Lett.}\ }\textbf {\bibinfo
  {volume} {96}},\ \bibinfo {pages} {207202} (\bibinfo {year}
  {2006})}\BibitemShut {NoStop}%
\bibitem [{\citenamefont {Ritz}\ \emph {et~al.}(2013)\citenamefont {Ritz},
  \citenamefont {Halder}, \citenamefont {Franz}, \citenamefont {Bauer},
  \citenamefont {Wagner}, \citenamefont {Bamler}, \citenamefont {Rosch},\ and\
  \citenamefont {Pfleiderer}}]{10}%
  \BibitemOpen
  \bibfield  {author} {\bibinfo {author} {\bibfnamefont {R.}~\bibnamefont
  {Ritz}}, \bibinfo {author} {\bibfnamefont {M.}~\bibnamefont {Halder}},
  \bibinfo {author} {\bibfnamefont {C.}~\bibnamefont {Franz}}, \bibinfo
  {author} {\bibfnamefont {A.}~\bibnamefont {Bauer}}, \bibinfo {author}
  {\bibfnamefont {M.}~\bibnamefont {Wagner}}, \bibinfo {author} {\bibfnamefont
  {R.}~\bibnamefont {Bamler}}, \bibinfo {author} {\bibfnamefont
  {A.}~\bibnamefont {Rosch}},\ and\ \bibinfo {author} {\bibfnamefont
  {C.}~\bibnamefont {Pfleiderer}},\ }\bibfield  {title} {\bibinfo {title}
  {Giant generic topological {H}all resistivity of {M}n{S}i under pressure},\
  }\href@noop {} {\bibfield  {journal} {\bibinfo  {journal} {Physical Review
  B}\ }\textbf {\bibinfo {volume} {87}},\ \bibinfo {pages} {134424} (\bibinfo
  {year} {2013})}\BibitemShut {NoStop}%
\bibitem [{\citenamefont {Binz}\ and\ \citenamefont {Vishwanath}(2008)}]{11}%
  \BibitemOpen
  \bibfield  {author} {\bibinfo {author} {\bibfnamefont {B.}~\bibnamefont
  {Binz}}\ and\ \bibinfo {author} {\bibfnamefont {A.}~\bibnamefont
  {Vishwanath}},\ }\bibfield  {title} {\bibinfo {title} {{Chirality induced
  anomalous-Hall effect in helical spin crystals}},\ }\href@noop {} {\bibfield
  {journal} {\bibinfo  {journal} {Phys. B Condens. Matter}\ }\textbf {\bibinfo
  {volume} {403}},\ \bibinfo {pages} {1336–1340} (\bibinfo {year}
  {2008})}\BibitemShut {NoStop}%
\bibitem [{\citenamefont {Büttner}\ \emph {et~al.}(2021)\citenamefont
  {Büttner}, \citenamefont {Pfau}, \citenamefont {Böttcher}, \citenamefont
  {Schneider},\ and\ \citenamefont {et~al.}}]{12}%
  \BibitemOpen
  \bibfield  {author} {\bibinfo {author} {\bibfnamefont {F.}~\bibnamefont
  {Büttner}}, \bibinfo {author} {\bibfnamefont {B.}~\bibnamefont {Pfau}},
  \bibinfo {author} {\bibfnamefont {M.}~\bibnamefont {Böttcher}}, \bibinfo
  {author} {\bibfnamefont {M.}~\bibnamefont {Schneider}},\ and\ \bibinfo
  {author} {\bibfnamefont {G.~M.}\ \bibnamefont {et~al.}},\ }\bibfield  {title}
  {\bibinfo {title} {{Observation of fluctuation-mediated picosecond nucleation
  of a topological phase}},\ }\href@noop {} {\bibfield  {journal} {\bibinfo
  {journal} {Nat. Mater.}\ }\textbf {\bibinfo {volume} {20}},\ \bibinfo {pages}
  {30–37} (\bibinfo {year} {2021})}\BibitemShut {NoStop}%
\bibitem [{\citenamefont {Shimojima}\ \emph {et~al.}(2021)\citenamefont
  {Shimojima}, \citenamefont {Nakamura}, \citenamefont {Yu}, \citenamefont
  {Karube}, \citenamefont {Taguchi}, \citenamefont {Tokura},\ and\
  \citenamefont {Ishizaka}}]{13}%
  \BibitemOpen
  \bibfield  {author} {\bibinfo {author} {\bibfnamefont {T.}~\bibnamefont
  {Shimojima}}, \bibinfo {author} {\bibfnamefont {A.}~\bibnamefont {Nakamura}},
  \bibinfo {author} {\bibfnamefont {X.}~\bibnamefont {Yu}}, \bibinfo {author}
  {\bibfnamefont {K.}~\bibnamefont {Karube}}, \bibinfo {author} {\bibfnamefont
  {Y.}~\bibnamefont {Taguchi}}, \bibinfo {author} {\bibfnamefont
  {Y.}~\bibnamefont {Tokura}},\ and\ \bibinfo {author} {\bibfnamefont
  {K.}~\bibnamefont {Ishizaka}},\ }\bibfield  {title} {\bibinfo {title}
  {Nano-to-micro spatiotemporal imaging of magnetic skyrmion's life cycle},\
  }\href {https://doi.org/10.1126/sciadv.abg1322} {\bibfield  {journal}
  {\bibinfo  {journal} {Science Advances}\ }\textbf {\bibinfo {volume} {7}},\
  \bibinfo {pages} {eabg1322} (\bibinfo {year} {2021})}\BibitemShut {NoStop}%
\bibitem [{\citenamefont {Pappas}\ \emph {et~al.}(2009)\citenamefont {Pappas},
  \citenamefont {Leli\`evre-Berna}, \citenamefont {Falus}, \citenamefont
  {Bentley}, \citenamefont {Moskvin}, \citenamefont {Grigoriev}, \citenamefont
  {Fouquet},\ and\ \citenamefont {Farago}}]{14}%
  \BibitemOpen
  \bibfield  {author} {\bibinfo {author} {\bibfnamefont {C.}~\bibnamefont
  {Pappas}}, \bibinfo {author} {\bibfnamefont {E.}~\bibnamefont
  {Leli\`evre-Berna}}, \bibinfo {author} {\bibfnamefont {P.}~\bibnamefont
  {Falus}}, \bibinfo {author} {\bibfnamefont {P.~M.}\ \bibnamefont {Bentley}},
  \bibinfo {author} {\bibfnamefont {E.}~\bibnamefont {Moskvin}}, \bibinfo
  {author} {\bibfnamefont {S.}~\bibnamefont {Grigoriev}}, \bibinfo {author}
  {\bibfnamefont {P.}~\bibnamefont {Fouquet}},\ and\ \bibinfo {author}
  {\bibfnamefont {B.}~\bibnamefont {Farago}},\ }\bibfield  {title} {\bibinfo
  {title} {Chiral paramagnetic skyrmion-like phase in mnsi},\ }\href
  {https://doi.org/10.1103/PhysRevLett.102.197202} {\bibfield  {journal}
  {\bibinfo  {journal} {Phys. Rev. Lett.}\ }\textbf {\bibinfo {volume} {102}},\
  \bibinfo {pages} {197202} (\bibinfo {year} {2009})}\BibitemShut {NoStop}%
\bibitem [{\citenamefont {Wang}\ \emph {et~al.}(2019)\citenamefont {Wang},
  \citenamefont {Daniels}, \citenamefont {Liao}, \citenamefont {Zhao},
  \citenamefont {Wang}, \citenamefont {Koster}, \citenamefont {Rijnders},
  \citenamefont {Chang}, \citenamefont {Xiao},\ and\ \citenamefont {Wu}}]{15}%
  \BibitemOpen
  \bibfield  {author} {\bibinfo {author} {\bibfnamefont {W.}~\bibnamefont
  {Wang}}, \bibinfo {author} {\bibfnamefont {M.~W.}\ \bibnamefont {Daniels}},
  \bibinfo {author} {\bibfnamefont {Z.}~\bibnamefont {Liao}}, \bibinfo {author}
  {\bibfnamefont {Y.}~\bibnamefont {Zhao}}, \bibinfo {author} {\bibfnamefont
  {J.}~\bibnamefont {Wang}}, \bibinfo {author} {\bibfnamefont {G.}~\bibnamefont
  {Koster}}, \bibinfo {author} {\bibfnamefont {G.}~\bibnamefont {Rijnders}},
  \bibinfo {author} {\bibfnamefont {C.-Z.}\ \bibnamefont {Chang}}, \bibinfo
  {author} {\bibfnamefont {D.}~\bibnamefont {Xiao}},\ and\ \bibinfo {author}
  {\bibfnamefont {W.}~\bibnamefont {Wu}},\ }\bibfield  {title} {\bibinfo
  {title} {{Spin chirality fluctuation in two-dimensional ferromagnets with
  perpendicular magnetic anisotropy}},\ }\href@noop {} {\bibfield  {journal}
  {\bibinfo  {journal} {Nat. Mater.}\ }\textbf {\bibinfo {volume} {18}},\
  \bibinfo {pages} {1054–1059} (\bibinfo {year} {2019})}\BibitemShut
  {NoStop}%
\bibitem [{\citenamefont {Ishizuka}\ and\ \citenamefont {Nagaosa}(2018)}]{16}%
  \BibitemOpen
  \bibfield  {author} {\bibinfo {author} {\bibfnamefont {H.}~\bibnamefont
  {Ishizuka}}\ and\ \bibinfo {author} {\bibfnamefont {N.}~\bibnamefont
  {Nagaosa}},\ }\bibfield  {title} {\bibinfo {title} {{Spin chirality induced
  skew scattering and anomalous Hall effect in chiral magnets}},\ }\href@noop
  {} {\bibfield  {journal} {\bibinfo  {journal} {Sci. Adv.}\ }\textbf {\bibinfo
  {volume} {4}},\ \bibinfo {pages} {eaap9962} (\bibinfo {year}
  {2018})}\BibitemShut {NoStop}%
\bibitem [{\citenamefont {Kolincio}\ \emph {et~al.}(2021)\citenamefont
  {Kolincio}, \citenamefont {Hirschberger}, \citenamefont {Masell},
  \citenamefont {Gao}, \citenamefont {Kikkawaa}, \citenamefont {Taguchi},
  \citenamefont {h.~Arima}, \citenamefont {Nagaosa},\ and\ \citenamefont
  {Tokura}}]{17}%
  \BibitemOpen
  \bibfield  {author} {\bibinfo {author} {\bibfnamefont {K.}~\bibnamefont
  {Kolincio}}, \bibinfo {author} {\bibfnamefont {M.}~\bibnamefont
  {Hirschberger}}, \bibinfo {author} {\bibfnamefont {J.}~\bibnamefont
  {Masell}}, \bibinfo {author} {\bibfnamefont {S.}~\bibnamefont {Gao}},
  \bibinfo {author} {\bibfnamefont {A.}~\bibnamefont {Kikkawaa}}, \bibinfo
  {author} {\bibfnamefont {Y.}~\bibnamefont {Taguchi}}, \bibinfo {author}
  {\bibfnamefont {T.}~\bibnamefont {h.~Arima}}, \bibinfo {author}
  {\bibfnamefont {N.}~\bibnamefont {Nagaosa}},\ and\ \bibinfo {author}
  {\bibfnamefont {Y.}~\bibnamefont {Tokura}},\ }\bibfield  {title} {\bibinfo
  {title} {{Large Hall and Nernst responses from thermally induced spin
  chirality in a spin-trimer ferromagnet}},\ }\href@noop {} {\bibfield
  {journal} {\bibinfo  {journal} {Proc. Natl. Acad. Sci.}\ }\textbf {\bibinfo
  {volume} {118}},\ \bibinfo {pages} {17} (\bibinfo {year} {2021})}\BibitemShut
  {NoStop}%
\bibitem [{\citenamefont {Sorn}\ \emph {et~al.}(2021)\citenamefont {Sorn},
  \citenamefont {Yang},\ and\ \citenamefont {Paramekanti}}]{18}%
  \BibitemOpen
  \bibfield  {author} {\bibinfo {author} {\bibfnamefont {S.}~\bibnamefont
  {Sorn}}, \bibinfo {author} {\bibfnamefont {L.}~\bibnamefont {Yang}},\ and\
  \bibinfo {author} {\bibfnamefont {A.}~\bibnamefont {Paramekanti}},\
  }\bibfield  {title} {\bibinfo {title} {{Resonant optical topological Hall
  conductivity from skyrmions}},\ }\href@noop {} {\bibfield  {journal}
  {\bibinfo  {journal} {Phys. Rev. B}\ }\textbf {\bibinfo {volume} {104}},\
  \bibinfo {pages} {134419} (\bibinfo {year} {2021})}\BibitemShut {NoStop}%
\bibitem [{\citenamefont {Bartram}\ \emph {et~al.}(2020)\citenamefont
  {Bartram}, \citenamefont {Sorn}, \citenamefont {Li}, \citenamefont {Hwangbo},
  \citenamefont {Shen}, \citenamefont {Frontini}, \citenamefont {He},
  \citenamefont {Yu}, \citenamefont {Paramekanti},\ and\ \citenamefont
  {Yang}}]{19}%
  \BibitemOpen
  \bibfield  {author} {\bibinfo {author} {\bibfnamefont {F.~M.}\ \bibnamefont
  {Bartram}}, \bibinfo {author} {\bibfnamefont {S.}~\bibnamefont {Sorn}},
  \bibinfo {author} {\bibfnamefont {Z.}~\bibnamefont {Li}}, \bibinfo {author}
  {\bibfnamefont {K.}~\bibnamefont {Hwangbo}}, \bibinfo {author} {\bibfnamefont
  {S.}~\bibnamefont {Shen}}, \bibinfo {author} {\bibfnamefont {F.}~\bibnamefont
  {Frontini}}, \bibinfo {author} {\bibfnamefont {L.}~\bibnamefont {He}},
  \bibinfo {author} {\bibfnamefont {P.}~\bibnamefont {Yu}}, \bibinfo {author}
  {\bibfnamefont {A.}~\bibnamefont {Paramekanti}},\ and\ \bibinfo {author}
  {\bibfnamefont {L.}~\bibnamefont {Yang}},\ }\bibfield  {title} {\bibinfo
  {title} {{Anomalous Kerr effect in SrRuO$_3$ thin films}},\ }\href@noop {}
  {\bibfield  {journal} {\bibinfo  {journal} {Phys. Rev. B}\ }\textbf {\bibinfo
  {volume} {102}},\ \bibinfo {pages} {140408} (\bibinfo {year}
  {2020})}\BibitemShut {NoStop}%
\bibitem [{\citenamefont {Isobe}\ \emph {et~al.}(2020)\citenamefont {Isobe},
  \citenamefont {Xu},\ and\ \citenamefont {Fu}}]{20}%
  \BibitemOpen
  \bibfield  {author} {\bibinfo {author} {\bibfnamefont {H.}~\bibnamefont
  {Isobe}}, \bibinfo {author} {\bibfnamefont {S.-Y.}\ \bibnamefont {Xu}},\ and\
  \bibinfo {author} {\bibfnamefont {L.}~\bibnamefont {Fu}},\ }\bibfield
  {title} {\bibinfo {title} {{High-frequency rectification via chiral Bloch
  electrons}},\ }\href@noop {} {\bibfield  {journal} {\bibinfo  {journal} {Sci.
  Adv.}\ }\textbf {\bibinfo {volume} {6}},\ \bibinfo {pages} {eaay2497}
  (\bibinfo {year} {2020})}\BibitemShut {NoStop}%
\bibitem [{\citenamefont {Lyanda-Geller}\ \emph {et~al.}(2001)\citenamefont
  {Lyanda-Geller}, \citenamefont {Chun}, \citenamefont {Salamon}, \citenamefont
  {Goldbart}, \citenamefont {Han}, \citenamefont {Tomioka}, \citenamefont
  {Asamitsu},\ and\ \citenamefont {Tokura}}]{21}%
  \BibitemOpen
  \bibfield  {author} {\bibinfo {author} {\bibfnamefont {Y.}~\bibnamefont
  {Lyanda-Geller}}, \bibinfo {author} {\bibfnamefont {S.~H.}\ \bibnamefont
  {Chun}}, \bibinfo {author} {\bibfnamefont {M.~B.}\ \bibnamefont {Salamon}},
  \bibinfo {author} {\bibfnamefont {P.~M.}\ \bibnamefont {Goldbart}}, \bibinfo
  {author} {\bibfnamefont {P.~D.}\ \bibnamefont {Han}}, \bibinfo {author}
  {\bibfnamefont {Y.}~\bibnamefont {Tomioka}}, \bibinfo {author} {\bibfnamefont
  {A.}~\bibnamefont {Asamitsu}},\ and\ \bibinfo {author} {\bibfnamefont
  {Y.}~\bibnamefont {Tokura}},\ }\bibfield  {title} {\bibinfo {title} {{Charge
  transport in manganites: Hopping conduction, the anomalous Hall effect, and
  universal scaling}},\ }\href@noop {} {\bibfield  {journal} {\bibinfo
  {journal} {Phys. Rev. B}\ }\textbf {\bibinfo {volume} {63}},\ \bibinfo
  {pages} {184426} (\bibinfo {year} {2001})}\BibitemShut {NoStop}%
\bibitem [{\citenamefont {Yanagihara}\ and\ \citenamefont
  {Salamon}(2002)}]{22}%
  \BibitemOpen
  \bibfield  {author} {\bibinfo {author} {\bibfnamefont {H.}~\bibnamefont
  {Yanagihara}}\ and\ \bibinfo {author} {\bibfnamefont {M.~B.}\ \bibnamefont
  {Salamon}},\ }\bibfield  {title} {\bibinfo {title} {{Skyrmion Strings and the
  Anomalous Hall Effect in CrO$_2$}},\ }\href@noop {} {\bibfield  {journal}
  {\bibinfo  {journal} {Phys. Rev. Lett.}\ }\textbf {\bibinfo {volume} {89}},\
  \bibinfo {pages} {187201} (\bibinfo {year} {2002})}\BibitemShut {NoStop}%
\bibitem [{\citenamefont {Chun}\ \emph {et~al.}(2000)\citenamefont {Chun},
  \citenamefont {Salamon}, \citenamefont {Lyanda-Geller}, \citenamefont
  {Goldbart},\ and\ \citenamefont {Han}}]{23}%
  \BibitemOpen
  \bibfield  {author} {\bibinfo {author} {\bibfnamefont {S.~H.}\ \bibnamefont
  {Chun}}, \bibinfo {author} {\bibfnamefont {M.~B.}\ \bibnamefont {Salamon}},
  \bibinfo {author} {\bibfnamefont {Y.}~\bibnamefont {Lyanda-Geller}}, \bibinfo
  {author} {\bibfnamefont {P.~M.}\ \bibnamefont {Goldbart}},\ and\ \bibinfo
  {author} {\bibfnamefont {P.~D.}\ \bibnamefont {Han}},\ }\bibfield  {title}
  {\bibinfo {title} {{Magnetotransport in Manganites and the Role of Quantal
  Phases: Theory and Experiment}},\ }\href@noop {} {\bibfield  {journal}
  {\bibinfo  {journal} {Phys. Rev. Lett.}\ }\textbf {\bibinfo {volume} {84}},\
  \bibinfo {pages} {757–760} (\bibinfo {year} {2000})}\BibitemShut {NoStop}%
\bibitem [{\citenamefont {Chandragiri}\ \emph {et~al.}(2016)\citenamefont
  {Chandragiri}, \citenamefont {Iyer},\ and\ \citenamefont
  {Sampathkumaran}}]{24}%
  \BibitemOpen
  \bibfield  {author} {\bibinfo {author} {\bibfnamefont {V.}~\bibnamefont
  {Chandragiri}}, \bibinfo {author} {\bibfnamefont {K.~K.}\ \bibnamefont
  {Iyer}},\ and\ \bibinfo {author} {\bibfnamefont {E.~V.}\ \bibnamefont
  {Sampathkumaran}},\ }\bibfield  {title} {\bibinfo {title} {{Magnetic behavior
  of Gd$_3$Ru$_4$Al$_{12}$, a layered compound with distorted kagomé net}},\
  }\href@noop {} {\bibfield  {journal} {\bibinfo  {journal} {J. Phys. Condens.
  Matter}\ }\textbf {\bibinfo {volume} {28}},\ \bibinfo {pages} {286002}
  (\bibinfo {year} {2016})}\BibitemShut {NoStop}%
\bibitem [{\citenamefont {Kotsanidis}\ \emph {et~al.}(1990)\citenamefont
  {Kotsanidis}, \citenamefont {Yakinthos},\ and\ \citenamefont
  {Gamari-Seale}}]{25}%
  \BibitemOpen
  \bibfield  {author} {\bibinfo {author} {\bibfnamefont {P.~A.}\ \bibnamefont
  {Kotsanidis}}, \bibinfo {author} {\bibfnamefont {J.~K.}\ \bibnamefont
  {Yakinthos}},\ and\ \bibinfo {author} {\bibfnamefont {E.}~\bibnamefont
  {Gamari-Seale}},\ }\bibfield  {title} {\bibinfo {title} {{Magnetic properties
  of the ternary rare earth silicides R$_2$PdSi$_3$ (R = Pr, Nd, Gd, Tb, Dy,
  Ho, Er, Tm and Y)}},\ }\href@noop {} {\bibfield  {journal} {\bibinfo
  {journal} {J. Magn. Magn. Mater.}\ }\textbf {\bibinfo {volume} {87}},\
  \bibinfo {pages} {199–204} (\bibinfo {year} {1990})}\BibitemShut {NoStop}%
\bibitem [{\citenamefont {Hirschberger}\ \emph {et~al.}(2019)\citenamefont
  {Hirschberger}, \citenamefont {Nakajima}, \citenamefont {Gao}, \citenamefont
  {Peng}, \citenamefont {Kikkawa}, \citenamefont {Kurumaji}, \citenamefont
  {Kriener}, \citenamefont {Yamasaki}, \citenamefont {Sagayama}, \citenamefont
  {Nakao}, \citenamefont {Ohishi}, \citenamefont {Kakurai}, \citenamefont
  {Taguchi}, \citenamefont {Yu}, \citenamefont {Arima},\ and\ \citenamefont
  {Tokura}}]{26}%
  \BibitemOpen
  \bibfield  {author} {\bibinfo {author} {\bibfnamefont {M.}~\bibnamefont
  {Hirschberger}}, \bibinfo {author} {\bibfnamefont {T.}~\bibnamefont
  {Nakajima}}, \bibinfo {author} {\bibfnamefont {S.}~\bibnamefont {Gao}},
  \bibinfo {author} {\bibfnamefont {L.}~\bibnamefont {Peng}}, \bibinfo {author}
  {\bibfnamefont {A.}~\bibnamefont {Kikkawa}}, \bibinfo {author} {\bibfnamefont
  {T.}~\bibnamefont {Kurumaji}}, \bibinfo {author} {\bibfnamefont
  {M.}~\bibnamefont {Kriener}}, \bibinfo {author} {\bibfnamefont
  {Y.}~\bibnamefont {Yamasaki}}, \bibinfo {author} {\bibfnamefont
  {H.}~\bibnamefont {Sagayama}}, \bibinfo {author} {\bibfnamefont
  {H.}~\bibnamefont {Nakao}}, \bibinfo {author} {\bibfnamefont
  {K.}~\bibnamefont {Ohishi}}, \bibinfo {author} {\bibfnamefont
  {K.}~\bibnamefont {Kakurai}}, \bibinfo {author} {\bibfnamefont
  {Y.}~\bibnamefont {Taguchi}}, \bibinfo {author} {\bibfnamefont
  {X.}~\bibnamefont {Yu}}, \bibinfo {author} {\bibfnamefont {T.-h.}\
  \bibnamefont {Arima}},\ and\ \bibinfo {author} {\bibfnamefont
  {Y.}~\bibnamefont {Tokura}},\ }\bibfield  {title} {\bibinfo {title} {Skyrmion
  phase and competing magnetic orders on a breathing kagom{\'e} lattice},\
  }\href@noop {} {\bibfield  {journal} {\bibinfo  {journal} {Nature
  Communications}\ }\textbf {\bibinfo {volume} {10}},\ \bibinfo {pages} {5831}
  (\bibinfo {year} {2019})}\BibitemShut {NoStop}%
\bibitem [{\citenamefont {Kurumaji}\ \emph {et~al.}(2019)\citenamefont
  {Kurumaji}, \citenamefont {Nakajima}, \citenamefont {Hirschberger},
  \citenamefont {Kikkawa}, \citenamefont {Yamasaki}, \citenamefont {Sagayama},
  \citenamefont {Nakao}, \citenamefont {Taguchi}, \citenamefont {Arima},\ and\
  \citenamefont {Tokura}}]{27}%
  \BibitemOpen
  \bibfield  {author} {\bibinfo {author} {\bibfnamefont {T.}~\bibnamefont
  {Kurumaji}}, \bibinfo {author} {\bibfnamefont {T.}~\bibnamefont {Nakajima}},
  \bibinfo {author} {\bibfnamefont {M.}~\bibnamefont {Hirschberger}}, \bibinfo
  {author} {\bibfnamefont {A.}~\bibnamefont {Kikkawa}}, \bibinfo {author}
  {\bibfnamefont {Y.}~\bibnamefont {Yamasaki}}, \bibinfo {author}
  {\bibfnamefont {H.}~\bibnamefont {Sagayama}}, \bibinfo {author}
  {\bibfnamefont {H.}~\bibnamefont {Nakao}}, \bibinfo {author} {\bibfnamefont
  {Y.}~\bibnamefont {Taguchi}}, \bibinfo {author} {\bibfnamefont {T.-h.}\
  \bibnamefont {Arima}},\ and\ \bibinfo {author} {\bibfnamefont
  {Y.}~\bibnamefont {Tokura}},\ }\bibfield  {title} {\bibinfo {title} {Skyrmion
  lattice with a giant topological {H}all effect in a frustrated
  triangular-lattice magnet},\ }\href@noop {} {\bibfield  {journal} {\bibinfo
  {journal} {Science}\ }\textbf {\bibinfo {volume} {365}},\ \bibinfo {pages}
  {914} (\bibinfo {year} {2019})}\BibitemShut {NoStop}%
\bibitem [{\citenamefont {Gladyshevskii}\ \emph {et~al.}(1993)\citenamefont
  {Gladyshevskii}, \citenamefont {Strusievicz}, \citenamefont {Cenzual},\ and\
  \citenamefont {Parthé}}]{28}%
  \BibitemOpen
  \bibfield  {author} {\bibinfo {author} {\bibfnamefont {R.~E.}\ \bibnamefont
  {Gladyshevskii}}, \bibinfo {author} {\bibfnamefont {O.~R.}\ \bibnamefont
  {Strusievicz}}, \bibinfo {author} {\bibfnamefont {K.}~\bibnamefont
  {Cenzual}},\ and\ \bibinfo {author} {\bibfnamefont {E.}~\bibnamefont
  {Parthé}},\ }\bibfield  {title} {\bibinfo {title} {{Structure of
  Gd$_3$Ru$_4$Al$_{12}$, a new member of the EuMg$_{5.2}$ structure family with
  minority-atom clusters}},\ }\href@noop {} {\bibfield  {journal} {\bibinfo
  {journal} {Acta Crystallogr. B}\ }\textbf {\bibinfo {volume} {49}},\ \bibinfo
  {pages} {474–478} (\bibinfo {year} {1993})}\BibitemShut {NoStop}%
\bibitem [{\citenamefont {Tang}\ \emph {et~al.}(2011)\citenamefont {Tang},
  \citenamefont {Frontzek}, \citenamefont {Dshemuchadse}, \citenamefont
  {Leisegang}, \citenamefont {Zschornak}, \citenamefont {Mietrach},
  \citenamefont {Hoffmann}, \citenamefont {L\"oser}, \citenamefont {Gemming},
  \citenamefont {Meyer},\ and\ \citenamefont {Loewenhaupt}}]{29}%
  \BibitemOpen
  \bibfield  {author} {\bibinfo {author} {\bibfnamefont {F.}~\bibnamefont
  {Tang}}, \bibinfo {author} {\bibfnamefont {M.}~\bibnamefont {Frontzek}},
  \bibinfo {author} {\bibfnamefont {J.}~\bibnamefont {Dshemuchadse}}, \bibinfo
  {author} {\bibfnamefont {T.}~\bibnamefont {Leisegang}}, \bibinfo {author}
  {\bibfnamefont {M.}~\bibnamefont {Zschornak}}, \bibinfo {author}
  {\bibfnamefont {R.}~\bibnamefont {Mietrach}}, \bibinfo {author}
  {\bibfnamefont {J.-U.}\ \bibnamefont {Hoffmann}}, \bibinfo {author}
  {\bibfnamefont {W.}~\bibnamefont {L\"oser}}, \bibinfo {author} {\bibfnamefont
  {S.}~\bibnamefont {Gemming}}, \bibinfo {author} {\bibfnamefont {D.~C.}\
  \bibnamefont {Meyer}},\ and\ \bibinfo {author} {\bibfnamefont
  {M.}~\bibnamefont {Loewenhaupt}},\ }\bibfield  {title} {\bibinfo {title}
  {{Crystallographic superstructure in $R_2$PdSi$_3$ compounds $R$ = rare
  earth)}},\ }\href {https://doi.org/10.1103/PhysRevB.84.104105} {\bibfield
  {journal} {\bibinfo  {journal} {Phys. Rev. B}\ }\textbf {\bibinfo {volume}
  {84}},\ \bibinfo {pages} {104105} (\bibinfo {year} {2011})}\BibitemShut
  {NoStop}%
\bibitem [{\citenamefont {Nakamura}\ \emph {et~al.}(2018)\citenamefont
  {Nakamura}, \citenamefont {Kabeya}, \citenamefont {Kobayashi}, \citenamefont
  {Araki}, \citenamefont {Katoh},\ and\ \citenamefont {Ochiai}}]{30}%
  \BibitemOpen
  \bibfield  {author} {\bibinfo {author} {\bibfnamefont {S.}~\bibnamefont
  {Nakamura}}, \bibinfo {author} {\bibfnamefont {N.}~\bibnamefont {Kabeya}},
  \bibinfo {author} {\bibfnamefont {M.}~\bibnamefont {Kobayashi}}, \bibinfo
  {author} {\bibfnamefont {K.}~\bibnamefont {Araki}}, \bibinfo {author}
  {\bibfnamefont {K.}~\bibnamefont {Katoh}},\ and\ \bibinfo {author}
  {\bibfnamefont {A.}~\bibnamefont {Ochiai}},\ }\bibfield  {title} {\bibinfo
  {title} {{Spin trimer formation in the metallic compound
  ${\mathrm{Gd}}_{3}{\mathrm{Ru}}_{4}{\mathrm{Al}}_{12}$ with a distorted
  kagome lattice structure}},\ }\href
  {https://doi.org/10.1103/PhysRevB.98.054410} {\bibfield  {journal} {\bibinfo
  {journal} {Phys. Rev. B}\ }\textbf {\bibinfo {volume} {98}},\ \bibinfo
  {pages} {054410} (\bibinfo {year} {2018})}\BibitemShut {NoStop}%
\bibitem [{\citenamefont {Han}\ and\ \citenamefont {et~al.}(2012)}]{31}%
  \BibitemOpen
  \bibfield  {author} {\bibinfo {author} {\bibfnamefont {T.-H.}\ \bibnamefont
  {Han}}\ and\ \bibinfo {author} {\bibnamefont {et~al.}},\ }\bibfield  {title}
  {\bibinfo {title} {{Fractionalized excitations in the spin-liquid state of a
  kagome-lattice antiferromagnet}},\ }\href@noop {} {\bibfield  {journal}
  {\bibinfo  {journal} {Nature}\ }\textbf {\bibinfo {volume} {492}},\ \bibinfo
  {pages} {406–410} (\bibinfo {year} {2012})}\BibitemShut {NoStop}%
\bibitem [{\citenamefont {Ohgushi}\ \emph {et~al.}(2000)\citenamefont
  {Ohgushi}, \citenamefont {Murakami},\ and\ \citenamefont {Nagaosa}}]{32}%
  \BibitemOpen
  \bibfield  {author} {\bibinfo {author} {\bibfnamefont {K.}~\bibnamefont
  {Ohgushi}}, \bibinfo {author} {\bibfnamefont {S.}~\bibnamefont {Murakami}},\
  and\ \bibinfo {author} {\bibfnamefont {N.}~\bibnamefont {Nagaosa}},\
  }\bibfield  {title} {\bibinfo {title} {{Spin anisotropy and quantum Hall
  effect in the kagomé lattice: Chiral spin state based on a ferromagnet}},\
  }\href@noop {} {\bibfield  {journal} {\bibinfo  {journal} {Phys. Rev. B}\
  }\textbf {\bibinfo {volume} {62}},\ \bibinfo {pages} {R6065–R6068}
  (\bibinfo {year} {2000})}\BibitemShut {NoStop}%
\bibitem [{\citenamefont {Katsura}\ \emph {et~al.}(2010)\citenamefont
  {Katsura}, \citenamefont {Nagaosa},\ and\ \citenamefont {Lee}}]{33}%
  \BibitemOpen
  \bibfield  {author} {\bibinfo {author} {\bibfnamefont {H.}~\bibnamefont
  {Katsura}}, \bibinfo {author} {\bibfnamefont {N.}~\bibnamefont {Nagaosa}},\
  and\ \bibinfo {author} {\bibfnamefont {P.~A.}\ \bibnamefont {Lee}},\
  }\bibfield  {title} {\bibinfo {title} {{Theory of the Thermal Hall Effect in
  Quantum Magnets}},\ }\href@noop {} {\bibfield  {journal} {\bibinfo  {journal}
  {Phys. Rev. Lett.}\ }\textbf {\bibinfo {volume} {104}},\ \bibinfo {pages}
  {066403} (\bibinfo {year} {2010})}\BibitemShut {NoStop}%
\bibitem [{\citenamefont {Chisnell}\ \emph {et~al.}(2015)\citenamefont
  {Chisnell}, \citenamefont {Helton}, \citenamefont {Freedman}, \citenamefont
  {Singh}, \citenamefont {Bewley}, \citenamefont {Nocera},\ and\ \citenamefont
  {Lee}}]{34}%
  \BibitemOpen
  \bibfield  {author} {\bibinfo {author} {\bibfnamefont {R.}~\bibnamefont
  {Chisnell}}, \bibinfo {author} {\bibfnamefont {J.~S.}\ \bibnamefont
  {Helton}}, \bibinfo {author} {\bibfnamefont {D.~E.}\ \bibnamefont
  {Freedman}}, \bibinfo {author} {\bibfnamefont {D.~K.}\ \bibnamefont {Singh}},
  \bibinfo {author} {\bibfnamefont {R.~I.}\ \bibnamefont {Bewley}}, \bibinfo
  {author} {\bibfnamefont {D.~G.}\ \bibnamefont {Nocera}},\ and\ \bibinfo
  {author} {\bibfnamefont {Y.~S.}\ \bibnamefont {Lee}},\ }\bibfield  {title}
  {\bibinfo {title} {Topological magnon bands in a kagome lattice
  ferromagnet},\ }\href {https://doi.org/10.1103/PhysRevLett.115.147201}
  {\bibfield  {journal} {\bibinfo  {journal} {Phys. Rev. Lett.}\ }\textbf
  {\bibinfo {volume} {115}},\ \bibinfo {pages} {147201} (\bibinfo {year}
  {2015})}\BibitemShut {NoStop}%
\bibitem [{\citenamefont {Kawamura}\ and\ \citenamefont
  {Miyashita}(1984)}]{35}%
  \BibitemOpen
  \bibfield  {author} {\bibinfo {author} {\bibfnamefont {H.}~\bibnamefont
  {Kawamura}}\ and\ \bibinfo {author} {\bibfnamefont {S.}~\bibnamefont
  {Miyashita}},\ }\bibfield  {title} {\bibinfo {title} {{Phase Transition of
  the Two-Dimensional Heisenberg Antiferromagnet on the Triangular Lattice}},\
  }\href@noop {} {\bibfield  {journal} {\bibinfo  {journal} {J. Phys. Soc.
  Jpn.}\ }\textbf {\bibinfo {volume} {53}},\ \bibinfo {pages} {4138–4154}
  (\bibinfo {year} {1984})}\BibitemShut {NoStop}%
\bibitem [{\citenamefont {Kawamura}\ \emph {et~al.}(2010)\citenamefont
  {Kawamura}, \citenamefont {Yamamoto},\ and\ \citenamefont {Okubo}}]{36}%
  \BibitemOpen
  \bibfield  {author} {\bibinfo {author} {\bibfnamefont {H.}~\bibnamefont
  {Kawamura}}, \bibinfo {author} {\bibfnamefont {A.}~\bibnamefont {Yamamoto}},\
  and\ \bibinfo {author} {\bibfnamefont {T.}~\bibnamefont {Okubo}},\ }\bibfield
   {title} {\bibinfo {title} {{$Z_2$-Vortex Ordering of the Triangular-Lattice
  Heisenberg Antiferromagnet}},\ }\href@noop {} {\bibfield  {journal} {\bibinfo
   {journal} {J. Phys. Soc. Jpn.}\ }\textbf {\bibinfo {volume} {79}},\ \bibinfo
  {pages} {023701} (\bibinfo {year} {2010})}\BibitemShut {NoStop}%
\bibitem [{\citenamefont {Lau}\ and\ \citenamefont {Dasgupta}(1989)}]{37}%
  \BibitemOpen
  \bibfield  {author} {\bibinfo {author} {\bibfnamefont {M.}~\bibnamefont
  {Lau}}\ and\ \bibinfo {author} {\bibfnamefont {C.}~\bibnamefont {Dasgupta}},\
  }\bibfield  {title} {\bibinfo {title} {{Numerical investigation of the role
  of topological defects in the three-dimensional Heisenberg transition}},\
  }\href@noop {} {\bibfield  {journal} {\bibinfo  {journal} {Phys. Rev. B}\
  }\textbf {\bibinfo {volume} {39}},\ \bibinfo {pages} {7212–7222} (\bibinfo
  {year} {1989})}\BibitemShut {NoStop}%
\bibitem [{\citenamefont {Lau}\ and\ \citenamefont {Dasgupta}(1988)}]{38}%
  \BibitemOpen
  \bibfield  {author} {\bibinfo {author} {\bibfnamefont {M.}~\bibnamefont
  {Lau}}\ and\ \bibinfo {author} {\bibfnamefont {C.}~\bibnamefont {Dasgupta}},\
  }\bibfield  {title} {\bibinfo {title} {{Role of topological defects in the
  phase transition of the three-dimensional Heisenberg model}},\ }\href@noop {}
  {\bibfield  {journal} {\bibinfo  {journal} {J. Phys. Math. Gen.}\ }\textbf
  {\bibinfo {volume} {21}},\ \bibinfo {pages} {L51–L57} (\bibinfo {year}
  {1988})}\BibitemShut {NoStop}%
\bibitem [{\citenamefont {Onoda}\ \emph {et~al.}(2006)\citenamefont {Onoda},
  \citenamefont {Sugimoto},\ and\ \citenamefont {Nagaosa}}]{39}%
  \BibitemOpen
  \bibfield  {author} {\bibinfo {author} {\bibfnamefont {S.}~\bibnamefont
  {Onoda}}, \bibinfo {author} {\bibfnamefont {N.}~\bibnamefont {Sugimoto}},\
  and\ \bibinfo {author} {\bibfnamefont {N.}~\bibnamefont {Nagaosa}},\
  }\bibfield  {title} {\bibinfo {title} {{Intrinsic Versus Extrinsic Anomalous
  Hall Effect in Ferromagnets}},\ }\href@noop {} {\bibfield  {journal}
  {\bibinfo  {journal} {Phys. Rev. Lett.}\ }\textbf {\bibinfo {volume} {97}},\
  \bibinfo {pages} {126602} (\bibinfo {year} {2006})}\BibitemShut {NoStop}%
\bibitem [{\citenamefont {Bogdanov}\ and\ \citenamefont
  {Yablonskii}(1989)}]{40}%
  \BibitemOpen
  \bibfield  {author} {\bibinfo {author} {\bibfnamefont {A.~N.}\ \bibnamefont
  {Bogdanov}}\ and\ \bibinfo {author} {\bibfnamefont {D.~A.}\ \bibnamefont
  {Yablonskii}},\ }\bibfield  {title} {\bibinfo {title} {{Thermodynamically
  stable ‘vortices’ in magnetically ordered crystals. The mixed state of
  magnets}},\ }\href@noop {} {\bibfield  {journal} {\bibinfo  {journal} {Sov.
  Phys. JETP}\ }\textbf {\bibinfo {volume} {68}},\ \bibinfo {pages} {101}
  (\bibinfo {year} {1989})}\BibitemShut {NoStop}%
\bibitem [{\citenamefont {Hirschberger}\ \emph {et~al.}(2020)\citenamefont
  {Hirschberger}, \citenamefont {Spitz}, \citenamefont {Nomoto}, \citenamefont
  {Kurumaji}, \citenamefont {Gao}, \citenamefont {Masell}, \citenamefont
  {Nakajima}, \citenamefont {Kikkawa}, \citenamefont {Yamasaki}, \citenamefont
  {Sagayama}, \citenamefont {Nakao}, \citenamefont {Taguchi}, \citenamefont
  {Arita}, \citenamefont {Arima},\ and\ \citenamefont {Tokura}}]{41}%
  \BibitemOpen
  \bibfield  {author} {\bibinfo {author} {\bibfnamefont {M.}~\bibnamefont
  {Hirschberger}}, \bibinfo {author} {\bibfnamefont {L.}~\bibnamefont {Spitz}},
  \bibinfo {author} {\bibfnamefont {T.}~\bibnamefont {Nomoto}}, \bibinfo
  {author} {\bibfnamefont {T.}~\bibnamefont {Kurumaji}}, \bibinfo {author}
  {\bibfnamefont {S.}~\bibnamefont {Gao}}, \bibinfo {author} {\bibfnamefont
  {J.}~\bibnamefont {Masell}}, \bibinfo {author} {\bibfnamefont
  {T.}~\bibnamefont {Nakajima}}, \bibinfo {author} {\bibfnamefont
  {A.}~\bibnamefont {Kikkawa}}, \bibinfo {author} {\bibfnamefont
  {Y.}~\bibnamefont {Yamasaki}}, \bibinfo {author} {\bibfnamefont
  {H.}~\bibnamefont {Sagayama}}, \bibinfo {author} {\bibfnamefont
  {H.}~\bibnamefont {Nakao}}, \bibinfo {author} {\bibfnamefont
  {Y.}~\bibnamefont {Taguchi}}, \bibinfo {author} {\bibfnamefont
  {R.}~\bibnamefont {Arita}}, \bibinfo {author} {\bibfnamefont {T.-H.}\
  \bibnamefont {Arima}},\ and\ \bibinfo {author} {\bibfnamefont
  {Y.}~\bibnamefont {Tokura}},\ }\bibfield  {title} {\bibinfo {title}
  {{Topological Nernst Effect of the Two-Dimensional Skyrmion Lattice}},\
  }\href@noop {} {\bibfield  {journal} {\bibinfo  {journal} {Phys. Rev. Lett.}\
  }\textbf {\bibinfo {volume} {125}},\ \bibinfo {pages} {076602} (\bibinfo
  {year} {2020})}\BibitemShut {NoStop}%
\bibitem [{\citenamefont {Ding}\ \emph {et~al.}(2019)\citenamefont {Ding},
  \citenamefont {Koo}, \citenamefont {Xu}, \citenamefont {Li}, \citenamefont
  {Lu}, \citenamefont {Zhao}, \citenamefont {Wang}, \citenamefont {Yin},
  \citenamefont {Lei}, \citenamefont {Yan}, \citenamefont {Zhu},\ and\
  \citenamefont {Behnia}}]{42}%
  \BibitemOpen
  \bibfield  {author} {\bibinfo {author} {\bibfnamefont {L.}~\bibnamefont
  {Ding}}, \bibinfo {author} {\bibfnamefont {J.}~\bibnamefont {Koo}}, \bibinfo
  {author} {\bibfnamefont {L.}~\bibnamefont {Xu}}, \bibinfo {author}
  {\bibfnamefont {X.}~\bibnamefont {Li}}, \bibinfo {author} {\bibfnamefont
  {X.}~\bibnamefont {Lu}}, \bibinfo {author} {\bibfnamefont {L.}~\bibnamefont
  {Zhao}}, \bibinfo {author} {\bibfnamefont {Q.}~\bibnamefont {Wang}}, \bibinfo
  {author} {\bibfnamefont {Q.}~\bibnamefont {Yin}}, \bibinfo {author}
  {\bibfnamefont {H.}~\bibnamefont {Lei}}, \bibinfo {author} {\bibfnamefont
  {B.}~\bibnamefont {Yan}}, \bibinfo {author} {\bibfnamefont {Z.}~\bibnamefont
  {Zhu}},\ and\ \bibinfo {author} {\bibfnamefont {K.}~\bibnamefont {Behnia}},\
  }\bibfield  {title} {\bibinfo {title} {{Intrinsic Anomalous Nernst Effect
  Amplified by Disorder in a Half-Metallic Semimetal}},\ }\href@noop {}
  {\bibfield  {journal} {\bibinfo  {journal} {Phys. Rev. X}\ }\textbf {\bibinfo
  {volume} {9}},\ \bibinfo {pages} {041061} (\bibinfo {year}
  {2019})}\BibitemShut {NoStop}%
\bibitem [{\citenamefont {Xu}\ \emph {et~al.}(2020{\natexlab{a}})\citenamefont
  {Xu}, \citenamefont {Li}, \citenamefont {Lu}, \citenamefont {Collignon},
  \citenamefont {Fu}, \citenamefont {Koo}, \citenamefont {Fauqué},
  \citenamefont {Yan}, \citenamefont {Zhu1},\ and\ \citenamefont
  {Behnia}}]{43}%
  \BibitemOpen
  \bibfield  {author} {\bibinfo {author} {\bibfnamefont {L.}~\bibnamefont
  {Xu}}, \bibinfo {author} {\bibfnamefont {X.}~\bibnamefont {Li}}, \bibinfo
  {author} {\bibfnamefont {X.}~\bibnamefont {Lu}}, \bibinfo {author}
  {\bibfnamefont {C.}~\bibnamefont {Collignon}}, \bibinfo {author}
  {\bibfnamefont {H.}~\bibnamefont {Fu}}, \bibinfo {author} {\bibfnamefont
  {J.}~\bibnamefont {Koo}}, \bibinfo {author} {\bibfnamefont {B.}~\bibnamefont
  {Fauqué}}, \bibinfo {author} {\bibfnamefont {B.}~\bibnamefont {Yan}},
  \bibinfo {author} {\bibfnamefont {Z.}~\bibnamefont {Zhu1}},\ and\ \bibinfo
  {author} {\bibfnamefont {K.}~\bibnamefont {Behnia}},\ }\bibfield  {title}
  {\bibinfo {title} {{Finite-temperature violation of the anomalous transverse
  Wiedemann-Franz law}},\ }\href@noop {} {\bibfield  {journal} {\bibinfo
  {journal} {Sci. Adv.}\ }\textbf {\bibinfo {volume} {6}},\ \bibinfo {pages}
  {eaaz3522} (\bibinfo {year} {2020}{\natexlab{a}})}\BibitemShut {NoStop}%
\bibitem [{\citenamefont {Xu}\ \emph {et~al.}(2020{\natexlab{b}})\citenamefont
  {Xu}, \citenamefont {Li}, \citenamefont {Ding}, \citenamefont {Chen},
  \citenamefont {Sakai}, \citenamefont {Fauqué}, \citenamefont {Nakatsuji},
  \citenamefont {Zhu},\ and\ \citenamefont {Behnia}}]{44}%
  \BibitemOpen
  \bibfield  {author} {\bibinfo {author} {\bibfnamefont {L.}~\bibnamefont
  {Xu}}, \bibinfo {author} {\bibfnamefont {X.}~\bibnamefont {Li}}, \bibinfo
  {author} {\bibfnamefont {L.}~\bibnamefont {Ding}}, \bibinfo {author}
  {\bibfnamefont {T.}~\bibnamefont {Chen}}, \bibinfo {author} {\bibfnamefont
  {A.}~\bibnamefont {Sakai}}, \bibinfo {author} {\bibfnamefont
  {B.}~\bibnamefont {Fauqué}}, \bibinfo {author} {\bibfnamefont
  {S.}~\bibnamefont {Nakatsuji}}, \bibinfo {author} {\bibfnamefont
  {Z.}~\bibnamefont {Zhu}},\ and\ \bibinfo {author} {\bibfnamefont
  {K.}~\bibnamefont {Behnia}},\ }\bibfield  {title} {\bibinfo {title}
  {{Anomalous transverse response of Co$_2$MnGa and universality of the
  room-temperature $\alpha^A_{ij}/\sigma_{ij}^A$ ratio across topological
  magnets}},\ }\href@noop {} {\bibfield  {journal} {\bibinfo  {journal} {Phys.
  Rev. B}\ }\textbf {\bibinfo {volume} {101}},\ \bibinfo {pages} {180404}
  (\bibinfo {year} {2020}{\natexlab{b}})}\BibitemShut {NoStop}%
\bibitem [{\citenamefont {Williams}(1979)}]{45}%
  \BibitemOpen
  \bibfield  {author} {\bibinfo {author} {\bibfnamefont {R.}~\bibnamefont
  {Williams}},\ }\href@noop {} {\emph {\bibinfo {title} {The Geometrical
  foundation of natural structure. A source book of design}}}\ (\bibinfo
  {publisher} {Dover Publications},\ \bibinfo {year} {1979})\BibitemShut
  {NoStop}%
\bibitem [{\citenamefont {Holden}(1971)}]{46}%
  \BibitemOpen
  \bibfield  {author} {\bibinfo {author} {\bibfnamefont {A.}~\bibnamefont
  {Holden}},\ }\href@noop {} {\emph {\bibinfo {title} {Shapes, Space, and
  Symmetry. Shapes, Space, and Symmetry}}}\ (\bibinfo  {publisher} {Columbia
  University Press},\ \bibinfo {year} {1971})\BibitemShut {NoStop}%
\end{thebibliography}%

\begin{figure}[b]
  \centering
	\includegraphics[clip, trim=0.cm 0.cm 0.cm 0.cm, width=1.0\linewidth]{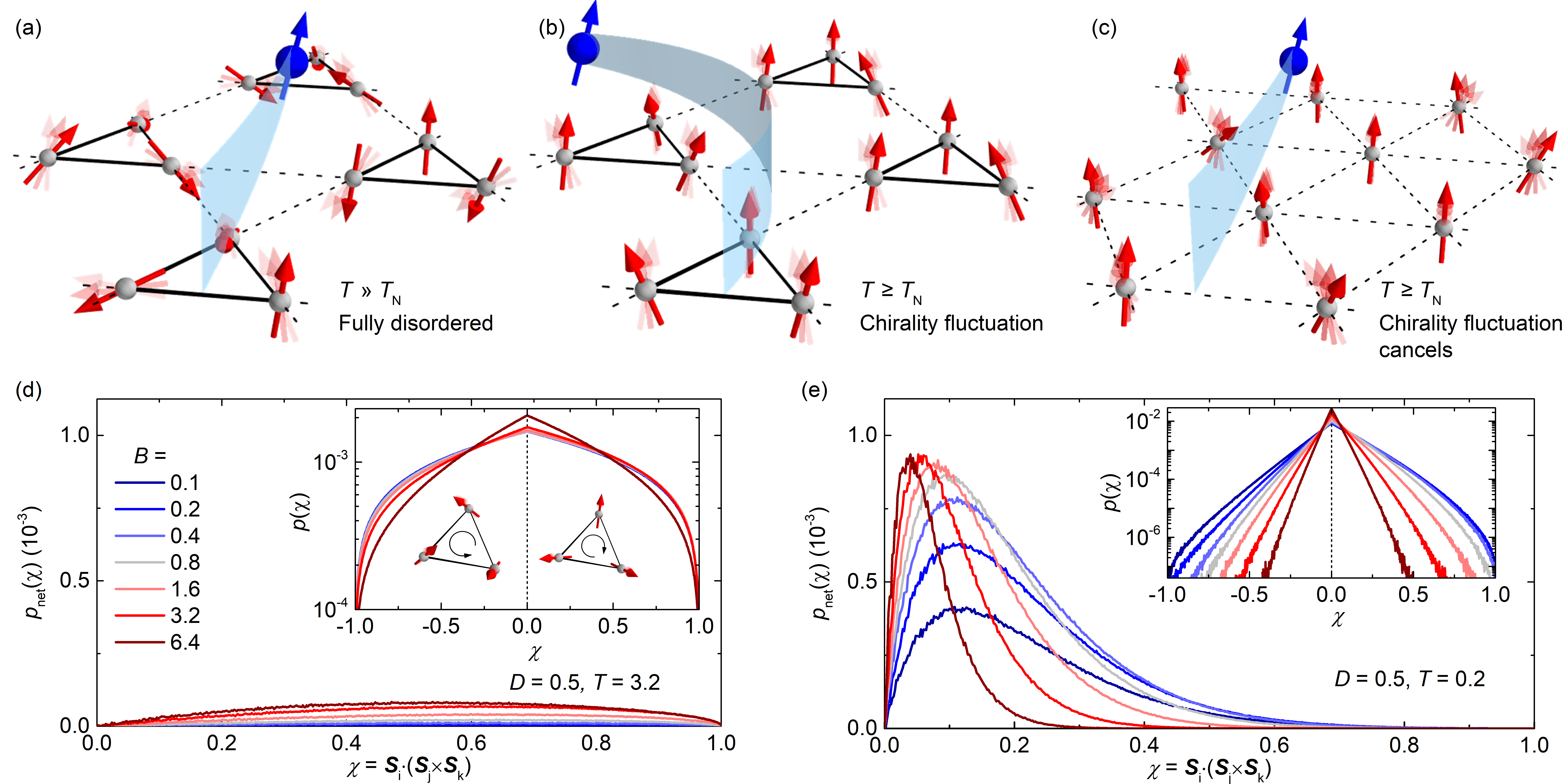}
  \caption{Spin fluctuations and spin chirality on a (breathing) kagomé lattice, and chirality cancellation a triangular lattice. (a,c) Thermal fluctuations of local spins (red) in the paramagnetic state, and their impact on moving conduction electrons (blue). We contrast (a), the fully disordered state with (b), a short-range correlated regime close to the critical temperature $T_N$ under the influence of chiral spin-spin interactions (Dzyaloshinskii-Moriya interaction $D$). (c) On a triangular lattice, no net chirality is produced and contributions from individual triangles cancel. (d,e) inset: Probability distribution $p(\chi)$ of the local scalar spin chirality $\chi$ on a single triangle. $B$ and $T$ are the external magnetic field and temperature respectively, expressed in the units of the nearest-neighbour exchange interaction $J = 1$. The inset cartoons illustrate exemplary spin configurations with negative and positive local chirality, respectively. (d,e) main panels: Antisymmetric part $p_\mathrm{net}(\chi)$ of $p(\chi)$. }
\label{fig:fig1}
\end{figure}

\begin{figure}[b]
  \centering
	\includegraphics[clip, trim=0.cm 0.cm 0.cm 0.cm, width=1.0\linewidth]{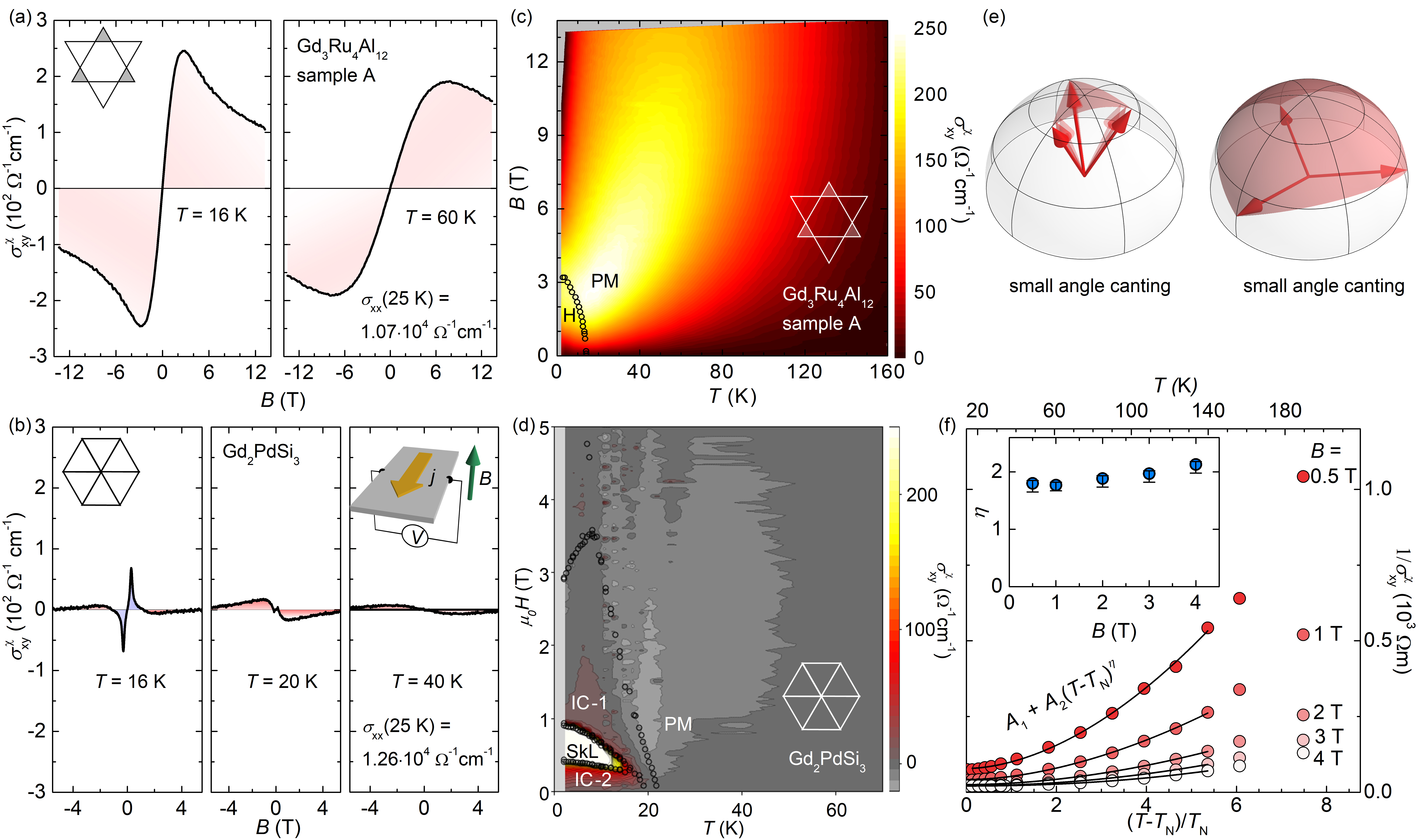}
  \caption{Chirality from spin fluctuations in transport experiments, and dependence on lattice geometry. (a,b) Chirality-driven Hall effect $\sigma_{xy}^\chi$ in kagomé-lattice Gd$_3$Ru$_4$Al$_{12}$ and in triangular-lattice Gd$_2$PdSi$_3$ at temperatures below and above the ordering temperature $T_N$. The red (blue) shaded areas indicate signals observed in the fluctuating regime (spin-ordered phase). Insets indicate the lattice type and measurement geometry. (c,d) Contour maps of $\sigma_{xy}^\chi$ in the plane spanned by magnetic field $B$ and $T$. The intensity scale is limited to $250\, \mathrm{\Omega^{-1}cm^{-1}}$ and magnetic phases are labelled as PM (paramagnetic), IC (incommensurate magnetic order), and SkL (skyrmion lattice). (e) Small and large angle canting of a spin triad. Specifically, small-angle fluctuations (left) compensate each other on trivial lattices, e.g. on the triangular grid. For a definition of trivial lattices, see Fig. 4. (f) Power-law fitting the temperature-induced decay of fluctuation-driven spin chirality. Inset shows exponent $\eta$ obtained from fits (black lines) to $\sigma_{xy}^\chi (T)\sim T^{–\eta}$ for data at various magnetic fields.}
\label{fig:fig2}
\end{figure}

\begin{figure}[b]
  \centering
	\includegraphics[clip, trim=0.cm 0.cm 0.cm 0.cm, width=1.0\linewidth]{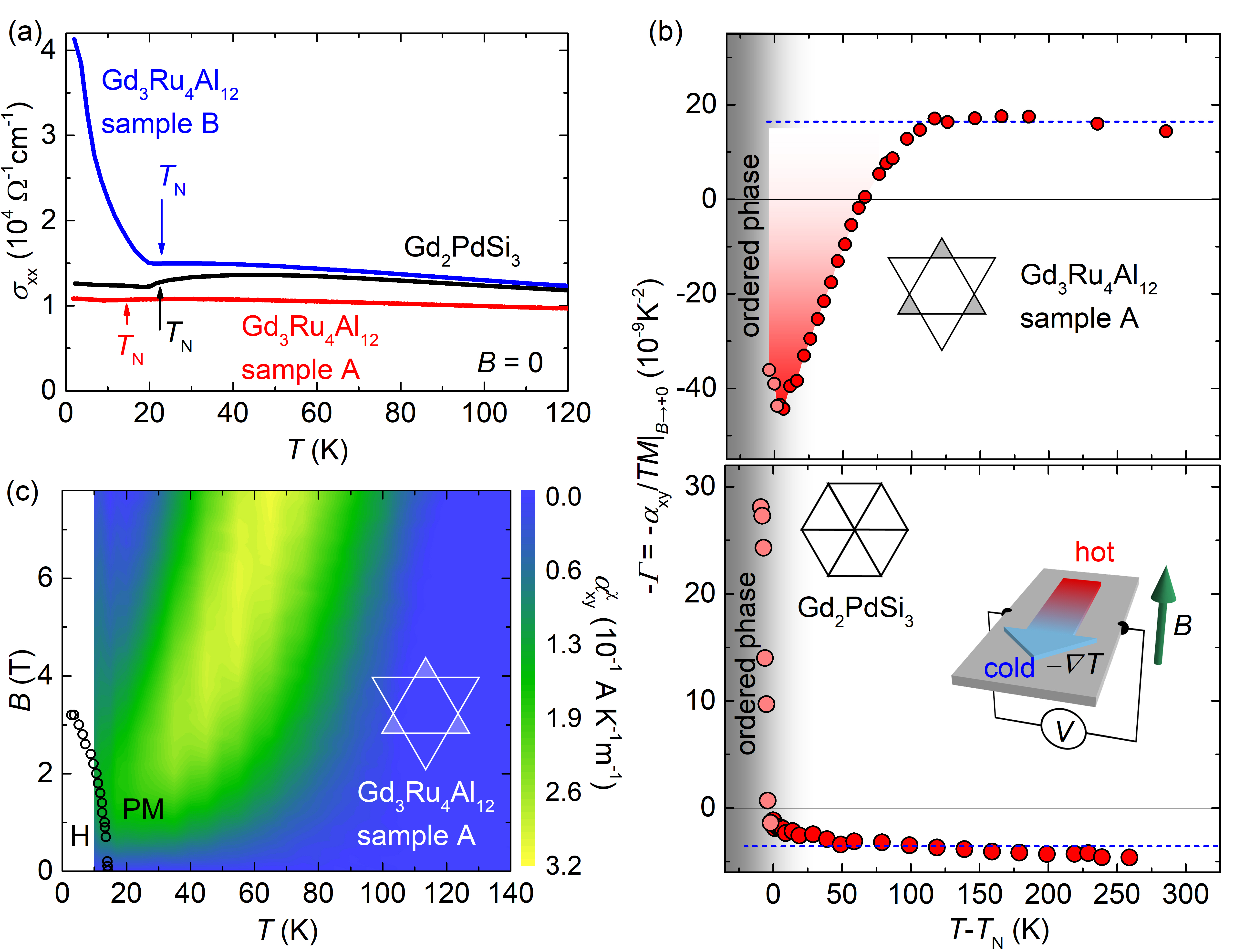}
  \caption{Nernst effect from spin fluctuations with chirality. (a) Longitudinal conductivity $\sigma_{xx}$, demonstrating comparable metallicity for Gd$_2$PdSi$_3$ (black colour) as well as samples A and B of Gd$_3$Ru$_4$Al$_{12}$ (red and blue colour). (b) Reduced low-field Nernst effect $\Gamma = \left.\alpha_{xy}/(TM)\right|_{B\rightarrow +0}$, i.e. total thermoelectric Nernst conductivity divided by temperature and magnetization, for kagomé lattice (upper) and triangular lattice (lower) compounds. Red and grey shading indicate the Nernst signal from thermal fluctuations and the long-range ordered regime, respectively. The blue dashed line is a measure of spin-orbit coupling driven anomalous Nernst effect. Scheme of experimental geometry used for Nernst effect measurements is displayed in the inset. See Methods for details. (c) Thermoelectric fluctuation-Nernst signal $\alpha_{xy}^\chi$ for the kagomé lattice in the temperature-magnetic field plane, after subtraction of a background. Black circles indicate the transition to long-range magnetic order. }
\label{fig:fig3}
\end{figure}

\begin{figure}[b]
  \centering
	\includegraphics[clip, trim=0.cm 0.cm 0.cm 0.cm, width=1.0\linewidth]{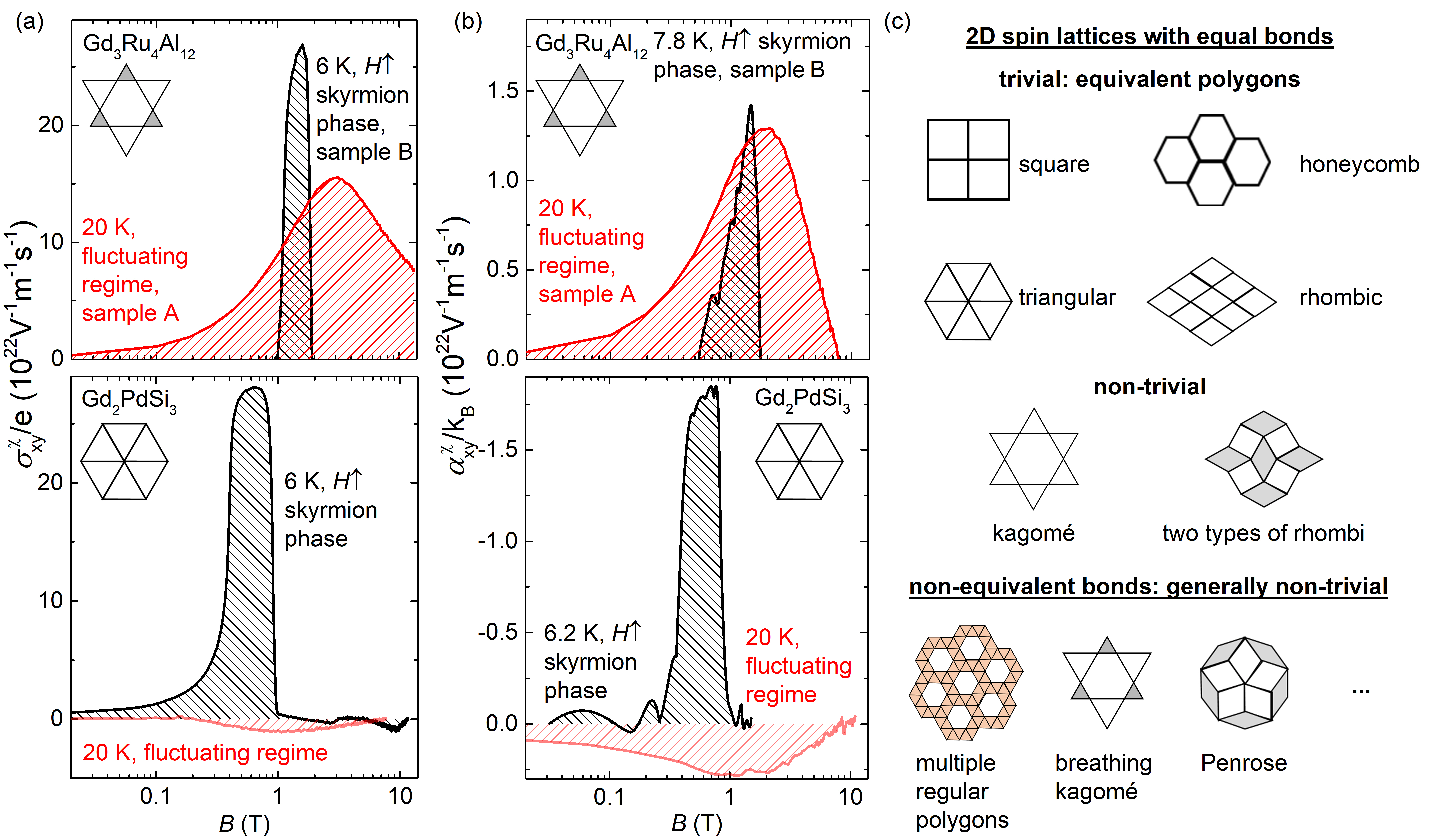}
  \caption{Comparing static and fluctuation-driven spin chirality on triangular and kagomé lattices. (a,b) Chirality-driven Hall conductivity $\sigma_{xy}^\chi$ (a) and Nernst conductivity $\alpha_{xy}^\chi$ (b) for kagomé and triangular lattices, with insets illustrating the lattice geometry. Black and red shading indicate signals in the static, long-range ordered skyrmion lattice phase and spin-dynamic (fluctuating) paramagnetic regimes, respectively. 
	(c) Classification of two-dimensional spin lattices as trivial and non-trivial, where trivial lattices lead to cancellation of spin chirality between neighbouring plaquettes. The three dots indicate a large number of additional nontrivial, two-dimensional lattices not depicted here.}
\label{fig:fig4}
\end{figure}

\end{document}